# Integration of Inter-Connectivity of Information System (i3) using Web Services

Aftab Ahmed Chandio*, Dingju Zhu, and Ali Hassan Sodhro, *Member, IAENG*

*Abstract*—**The growth and development of institutions day by day especially universities are increasing very large number of peoples including employees as well as students, working on different domain in distance between multiple integrated institutes, and degrading the overall efficiency of system that cannot facilitate better interoperability and automated integration. In many universities, exist system does not have yet set of connections, in spite of the fact they have different systems running on different platforms in different administrative units/departments. University of Sindh (UoS) is the one of them. In this paper we discuss about Integration of Inter-Connectivity of Information System (i3) in UoS based on Service Oriented Architecture (SOA) with web services. System i3 can monitor and exchange student's information in support of verification along heterogeneous and decentralized nature, and provide capability of interoperability in the place of already deployed system in UoS using different Languages as well as Databases.**

## I. INTRODUCTION

Each year in educational sectors are increasing huge number of peoples including students and employees, which has forced us to enhance the exist system. The existing deployed system cannot create automatics interconnection between departments when the departments need to exchange information and thus some delay created. They have only facilitated by manually communication due to some strong limitations. For example, every department works on different platform with different programming languages and different Databases. Therefore these departments need to integrate with each other to communicate automatically. The results of integration give facility of sharing and exchanging

Manuscript received December 12, 2011; revised January 18, 2012. This work was Supported by National Natural Science Foundation of China (grant no. 61105133), Shenzhen public technical service platform (grant no. ShenFaGai(2010)1324), and Shenzhen Internet Foundation (grant no. ShenFaGai(2011)1673).

Aftab Ahmed Chandio is a PhD student with Shenzhen Institutes of Advanced Technology, Chinese Academy of Science (SIAT-CAS), Graduate University of Chinese Academy of Sciences Beijing, China and He is also Lecturer on study leave with Institute of Mathematics & Computer Science, University of Sindh, Jamshoro Pakistan. (Phone: +8613244762252; e-mail: aftabac@ siat.ac.cn).

Dingju Zhu is an associate researcher with Shenzhen Institutes of Advanced Technology, Chinese Academy of Science (SIAT-CAS), Shenzhen, China. (Phone: +86 13316588865; e-mail: dj.zhu@siat.ac.cn).

Ali Hassan Sodhro is a PhD student with Shenzhen Institutes of Advanced Technology, Chinese Academy of Science (SIAT-CAS), Graduate University of Chinese Academy of Sciences Beijing, China. (e-mail: ali.hassan@siat.ac.cn).

information to different departments in educational sector. For example, university system is enclosed for communication with the extended campuses out of boundary, library department, student information department and university management department [1] [2]. The stimulation of new and rapid technological changes globally requires frequent upgrades and changes for such type of system, which is raising a finger for Service Oriented Architecture (SOA) with heterogeneous and decentralization manners. SOA can facilitate group of web services over the network, and provide the way of making enterprise system with distributed computers, databases and connections. This group of web services is the heart of SOA architecture, which plays a key role in SOA architecture and provides the business logic as services for heterogeneous environment [11]. Without replacing, deployed system can be extended by new system and application through integration or connectivity between them with the help of web services [3]. We all know that web service techniques are helpful to quickness and security, ability of interoperability, integration, platform and language independent. This paper illustrate system i3 in UoS, it can solve above problems specially in No-Dues verification from different departments, which are only needed at the time of final certificate issued by higher and confidential authority of university. This paper explains as follow in Section II: *Literature review*, Section III: *Basic overview of system i3*, Section IV: *How system i3 works and its solution*, Section V: *Related work* and last Section IV: *Conclusion*.

## II. LITERATURE REVIEW

Before we go ahead, first we have to look on basic principles of distributed computing. The suitable definition of distributed system "is a collection of independent computers that appears to its users as a single coherent system as the aspect deal with software" articulated by Andrew S. Tanenbaum [4]. A single program as a service can be run on network between multiple computers simultaneously; this way is only describing basic nature of distributed computing.

In 1980 decade, program-to-program was most fashionable work in the concept of distributed computing approach; many companies had been worked on this approach like System Application Architecture (SAA) by IBM working on own LU 6.2 API and PU 2.1 connection protocol, Remote Procedure Calls (RPCs) by Apollo, Distributed Network Architecture (DNA) by Microsoft, CORBA by Object Management Group and DCE by Open software foundation. Above approaches has strong limitation especially in data format, API and network environment, with hard-programming and limited types of applications. However web service is another distributed computing approach, which can support easy programming to find the distributed services request/response automatically and allow process communication in different APIs [5]. Web service is a technological backbone of Service Oriented Architecture (SOA). SOA can provide the way of design, development, deployment and management of business logic (i.e. functions, messages, services) within the Internet. The goal of SOA services can be scalable, flexible and loose coupled [6].

Web services can be provide standard-based way for best realizing SOA. There are three major roles that can explain the life cycle of web service in figure 1. *1. Client 2.Broker* and *3.Server*. The *client* can send request to broker for using service. A *broker* is performing intermediate role between service requester and provider, and it has service registry/directory. A *server* is provider of actual services. Simple Object Access Protocol (SOAP), eXtensible Markup Language (XML), Universal Description, Discovery and Integration (UDDI) and Web Service Description Language (WSDL) are the basic elements of web services. Request and response of service are circulating in the form of XML between three roles with the help of SOAP, and SOAP have three natures (find, publish and bind) for a complete process in life cycle of web services [7]. SOAP is an XML-based protocol to let applications exchange information over HTTP most used Internet protocol. Cross platform independency, language independency and interoperability are facilities and mechanisms of SOAP. UDDI is a directory for storing information of WSDL. WSDL is used to self-describe and locate the web services [12] [13].

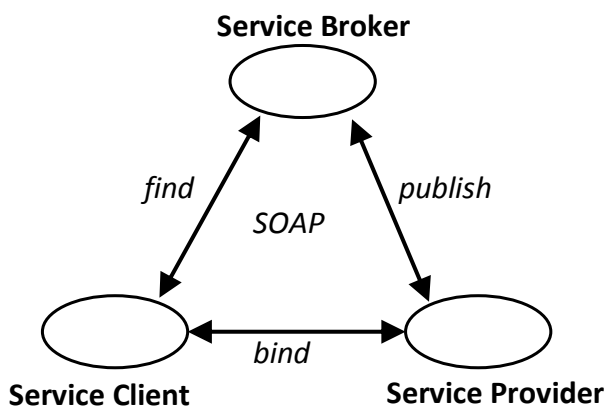

Fig. 1. Major roles of web services [7] [10]

## III. BASIC OVERVIEW OF SYSTEM I3

System i3 has major functions for sharing and exchanging all kind of information mostly associated with student in different departments, and thus each department can easily and quickly find the status of student for required results. In this paper we have solved many problems frequently occurred in university information system. Previous deployed system has different types of software, platform and database in departments. The integration and verification both are biggest problems of them. We have integrated all departments without replacing any deployed system. A verification process occurs at the time of issuing certificate needed by student and university. Student will be issued a degree certificate from higher authority of university after passing required examination for degree program, and before it, students has to clear No-Dues from all different departments of university. The i3 system gives facility that higher authority of examination in university can easily and quickly get a status report of student from all departments at the first hand and at a time.

The information in system i3 can be dividing into five basic levels. Figure 2 shows the relationships among five basic levels of information. Basic Student Information: this is the primary information of student like personal information (id, first name, last name, address, contact number and etc) and educational information (faculty, department, degree program, Graduation batch year and etc) and it can flow between all departments. Library Information: it is the information about books (id, isbn, title, author, publisher, year and etc) and students (personal and educational), and it has some kind of special report about students, whose have ordered the books. Hostel Information: it includes room and student information. It has also some report of special students who have allotted the room. Campuses Information: each campus need to store student information (personal and educational). There are many campuses located in distance. Campuses are academic extension of university. Examination Information: Department can manage all examination processes like storing students' examination records for issue final certificates. There are a lot of secrete information exchanged from other departments in verification process working on business logic from basic student, library to hostel information.

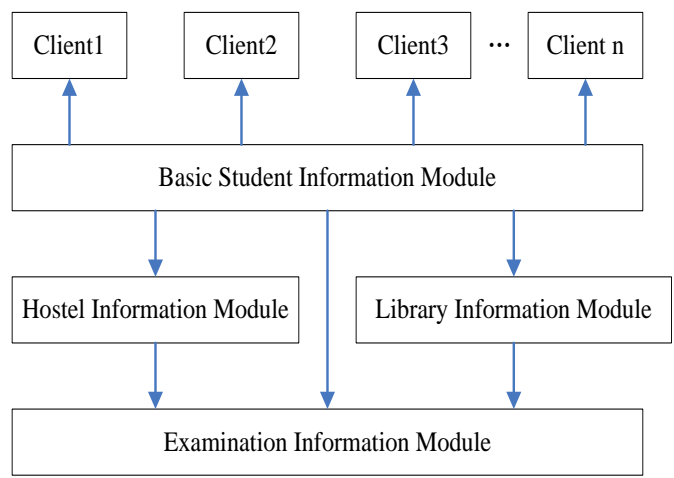

Fig. 2. Basic levels of information in system i3

Without system i3, these departments cannot exchange data automatically rather they can communicate manually. For example, all department need to access basic student information as common and one of them examination department need to use information from many department like library and hostel information. In other words this kind of information is called as service because each department need to access and use the information, which is clear mentioned in figure 2. Integration is the best solution for such type of enterprise system. The integration of knowledge and information process will assist in bringing up some standards for information collection, dissemination and management and also some other standards cataloguing, storing and retrieval of library data [1].

## IV. HOW SYSTEM i3 WORKS AND ITS SOLUTION

System i3 in UoS has deployed many web services without any costly or hard programming. Figure 5 explains the life cycle of system i3 using web services. System i3 has summarized in five modules, and sometimes modules can be increased more than five which already shown in figure 2. In this Section of paper, we describe all modules in brief. Administration Management Information System (AMIS), Library Management Information System (LMIS), Hostel Management Information System (HMIS) and Examination Management Information System (EMIS) are the major modules in system i3.

- AMIS plays a fundamental role in system i3, which can store basic student information like personal and educational information. AMIS provides information in the way of web service for reuse in each department. The information in AMIS mainly used in the form of web service by campuses for student registration in own database, and also it can be used by LMIS, HMIS and EMIS as web service. The example in figure 3 shows the page of student registration in LMIS module using AMIS web services.

- LMIS can be used to manage book information as well as student, and it accesses service from AMIS. This module stores status of student who might have ordered for books, which is called defaulter student information report that will be used by EMIS as web services.

- Likely LMIS, HMIS is accessing service from AMIS for student's registration and storing room information. It has also special information report about room allotted to student, which is the service mostly required by EMIS.

- The number of Campus information system may increase in future for new extension of university that can also use AMIS service for student registration in the campus system.

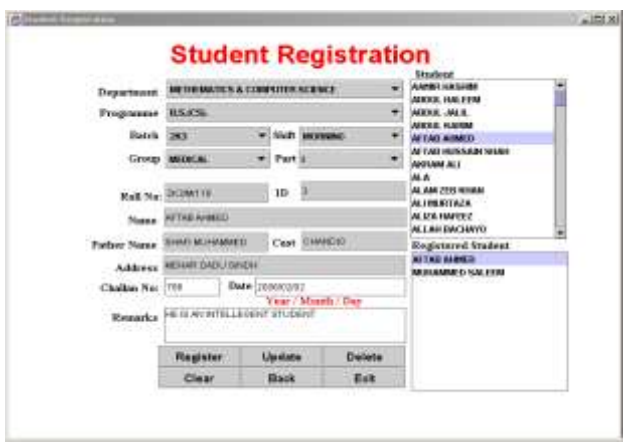

Fig. 3 The Student Registration page in LMIS using AMIS web services

- EMIS provides service in the last phase when students want to get certificates from higher authority of university after completing theirs degree program. In this stage EMIS application need to access all services even special reports from other departments at the same time. The certificate page of EMIS is shown in figure 4. After selecting student information using AMIS web service, the authority person can press verification button to find student's status, it is using web services provided by LMIS and HMIS.

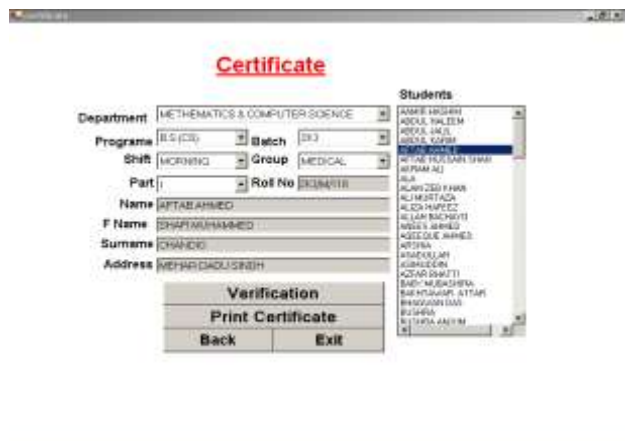

Fig. 4 The Certificate page in EMIS using web services (AMIS, HMIS, and LMIS)

System i3 in UoS has been developed based on the project of Apache software foundation for web services implementation called Apache eXtensible Interaction System (Axis), which is an open source SOAP engine [14]. We have implemented the SOAP in Axis server only programming with Java but it also can be programmed for service with C/C++. For deep understanding system i3, we should look on the life cycle of system i3 in figure 5, it shows that web service approach is making self-describable service and combining internal and external services to make up SOA.

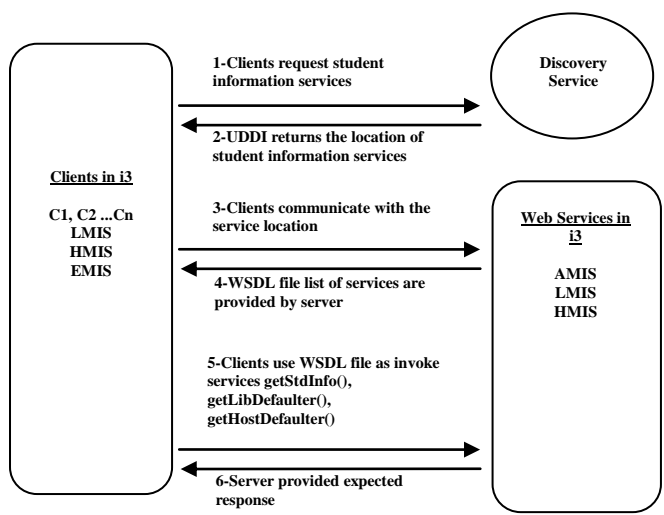

Fig. 5 The life cycle of system i3 using web services

Service requesters such as Client, LMIS, HMIS, EMIS request to access student's information from AMIS services, and one of them EMIS requests for access services from many providers like LMIS, HMIS and AMIS at the same time to monitor student's status for verification purpose. Service requesters find the location of service, which is registered in discovery service and was published by web service provider. WSDL file has a list of all services and way of its usage. Then clients can make a connection to the location of service according to its defined way, and they can invoke functions for exchanging information through the XML based message by SOAP.

When we deploy any service in Axis engine, we need to use of deployment descriptor which is pure XML-based, it called web service deployment descriptor (WSDD) [15]. In WSDD, we write many nodes, like <deployment> node for describing services in <service name> node, and node for theirs methods / functions, and <beanMapping> node for describe bean of their service. Axis engine can undeploy any service by using <undeployment> node. In below WSDD XML-based file 1, we have created three services for deployment in system i3, *AdmissionDataBaseManager*, *LibraryDataBaseManager* and *HostelDataBase-Manager.* And they have collection of many different methods and related records.

**WSDD XML-based file 1:**

```
<deployment xmlns="http://xml.apache.org/axis/wsdd/"
        xmlns:java="http://xml.apache.org/axis/wsdd/providers/java" >
  <handler name="print" type="java:LogHandler"/>

  <service name="AdmissionDataBaseManagerService" provider="java:RPC">
    <requestFlow>
      <handler type="print"/>
    </requestFlow>
    <beanMapping qname="myNS:StudentRecord"
        xmlns:myNS="urn:BeanService" languageSpecificType="java:StudentRecord"/>
    <beanMapping qname="myNS:DepartmentRecord"
        xmlns:myNS="urn:BeanService" languageSpecificType="java:DepartmentRecord"/>
    <beanMapping qname="myNS:ProgrammeRecord"
        xmlns:myNS="urn:BeanService"  languageSpecificType="java:ProgrammeRecord"/>
    <beanMapping qname="myNS:ListItem"
        xmlns:myNS="urn:BeanService" languageSpecificType="java:ListItem"/>
  </service>

  <service name="LibraryDataBaseManagerService" provider="java:RPC">
    <requestFlow>
      <handler type="print"/>
    </requestFlow>
    <beanMapping qname="myNS:LibraryStudentRecord"
        xmlns:myNS="urn:BeanService"  languageSpecificType="java:LibraryStudentRecord"/>
  </service>

  <service name="HostelDataBaseManagerService" provider="java:RPC">
    <requestFlow>
      <handler type="print"/>
    </requestFlow>
    <beanMapping qname="myNS:HostelStudentRecord"
xmlns:myNS="urn:BeanService"  languageSpecificType="java:HostelStudentRecord"/>
  </service>
    </deployment>
```

System i3 can allow similar and dissimilar applications to communicate and exchange data between each other through the XML based message by SOAP. We have deployed above three services in Apache Axis SOAP engine. It allows to every modules of system i3 can work on platform independent as well as language independent and store information on their own databases. For example, in system i3 LMIS works in Java platform with SQL server database, AMIS in Java platform with mySql database, and EMIS in .Net platform.

## V. RELATED WORK

Many problems and solutions have been discussed in [2], [5], [7], [8], [9], [10], [11] and scholars found the solution of problems through implementation of web services.

Alkhanak [2] addressed in his paper, SOAPGS system is created for University of Malaysia, it can facilitate to postgraduate students with many services for reducing cost and time in their daily academic life when students need to communicate to other departments.

Z. Panian [3] expressed some benefits of integration in an enterprise system. The integration process gives forward-looking and flexibility way, where organizations can achieve and view to data on the spot, and perform operations in real-time.

L. Li [7] studied on hardware platform with help of web services. The author described some addition code is needed in both side Java mobile phone and the service side for integration of mobile phone-oriented application of web services.

K.Liang [8] discussed about some factors of university system need to integration between their applications. There are so many factors can effect to management of universities because of shortage of funds and some technical problems for example all departments of university working independently, lake of standard planning, different departments have different levels of information technology, "Information isolated island" prominent and decision making support system.

A. Shaikh [9] discussed in his paper that web services is solution of increasing problems in distributed telemedicine system during data integration, vendor lock-in and interoperability.

Talis SOA architecture [11] provided Talis adaptor which contains group of services support heterogeneous behavior for different sources.

## VI. CONCLUSION

In this paper, we have discussed about our designed system i3 for UoS. From system i3, we can see that web service is a good solution for university SOA infrastructure. We have integrated many systems of UoS in one framework, without system i3 they could not work automatically with frequent intercommunication. System i3 has also solved another major problem of examination department in UoS. The problem was raised during verification of student from all departments for issuing final certificate, and it created more complicate when all departments have different types of platforms and databases. System i3 can not only implement in UoS, but also can be fit in other institutes where they have occurred similar problems.